\documentclass[pra,aps,twocolumn,showpacs]{revtex4}
\usepackage{times,epsfig,amssymb,graphicx,amsfonts,amsmath}
\newcommand{\ket}[1]{|#1\rangle}
\newcommand{\bra}[1]{\langle #1|}

\begin{document}

\title{Correlation evolution and monogamy of two geometric quantum discords in multipartite systems}

\author{Yan-Kui Bai$^{1,2}$}
\email{ykbai@semi.ac.cn}
\author{Ting-Ting Zhang$^{2}$}
\author{Li-Tao Wang$^{2}$}
\author{Z. D. Wang$^{1}$}
\email{zwang@hku.hk}

\affiliation{$^1$ Department of Physics and Center of Theoretical and Computational
Physics, The University of Hong Kong, Pokfulam Road, Hong Kong, China\\
$^2$ College of Physical Science and Information Engineering and Hebei Advance Thin
Films Laboratory, Hebei Normal University, Shijiazhuang, Hebei 050024, China}

\begin{abstract}
We explore two different geometric quantum discords defined respectively {\it via} the trace norm
(GQD-1) and Hilbert-Schmidt norm (GQD-2) in multipartite systems. A rigorous hierarchy relation is
revealed for the two GQDs in a class of symmetric two-qubit $X$-shape states. For multiqubit pure
states, it is found that both GQDs are related to the entanglement concurrence, with the hierarchy
relation being saturated. Furthermore, we look into a four-partite dynamical system
consisting of two cavities interacting with independent reservoirs. It is found that the GQD-2 can
exhibit various sudden change behaviours, while the GQD-1 only evolves asymptotically, with the two
GQDs exhibiting different monogamous properties.
\end{abstract}


\maketitle

\section{Introduction}

Quantum correlation beyond entanglement is a typical characteristic of quantumness, which has attracted
a lot of attention for the last decade~\cite{modi12rmp,datta13ijmpb,fran13ijmpb,jsxu13nc}. While quantum discord
\cite{ollivier01prl,vedral01jpa} is a kind of basic measure for bipartite quantum correlations,  it is
difficult to evaluate even for two-qubit states. Daki\'{c} \emph{et al} introduced
a geometric quantum discord {\it via} Hilbert-Schmidt norm (GQD-2), which is expressed as~\cite{dakic10prl}
\begin{equation}\label{1}
D_{g2}(\rho_{AB})=2\mbox{min}_{\chi\in \Omega_0}||\rho_{AB}-\chi_{AB}||_{2}^2,
\end{equation}
where the minimum runs over all the set $\Omega_{0}$ of zero-discord states
$\chi_{AB}=\sum_i\ket{i}\bra{i}_A\otimes \rho_{B}^i$, and the geometric distance is quantified by the
square norm $||X-Y||_2^2=\mbox{Tr}(X-Y)^2$. The GQD-2 is effortlessly computable
in many interesting cases \cite{luo10pra,gir12prl,rana12pra}, especially for two-qubit
mixed states. Moreover, it is recently shown that this measure is related to the fidelity of remote
state preparation \cite{dakic12np,gio13pra,hor13arx}.

Unfortunately, the GQD-2 may increase under local operations on the unmeasured party
\cite{xhu13pra, tuf12pra,pia12pra}, which is different from that of quantum discord and thus is argued
to be a drawback. To circumvent this problem, some other geometric discords may need to be introduced, such as
the geometric quantum discord based on the trace norm (GQD-1)~\cite{deb12pra,pau13pra,aar13pra,aar13njp}
\begin{equation}\label{2}
D_{g1}(\rho_{AB})=\mbox{min}_{\chi\in \Omega_0}||\rho_{AB}-\chi_{AB}||_{1},
\end{equation}
where the trace norm is $||X||_1=\mbox{Tr}\sqrt{X^{\dagger}X}$.  The GQD-1 has an operational
interpretation that it is in correspondence with the one-shot state distinguishability
between two states \cite{gil05pra}.
However, the computability of GQD-1 is weaker than that of GQD-2,
and the analytical formula is available only for some specific classes of two-qubit states
\cite{pau13pra,pau13epl,nak13pra,cic14njp}.

For two-qubit Bell-diagonal states, it was identified both theoretically and experimentally that
the GQD-1 can exhibit the freezing and sudden change behaviors under local noise environments
\cite{aar13pra,mon13pra,aar13njp,sil13prl,pau13prl}. In particular, in the dynamical procedures
preserving Bell-diagonal form, the two GQDs obey the following relation~\cite{pau13pra}
\begin{equation}\label{3}
D_{g1}^2[\rho^{BD}(t)]\geq D_{g2}[\rho^{BD}(t)].
\end{equation}
For the dynamical evolution beyond Bell-diagonal form, the relation between two GQDs may need a further
investigation. In particular, for multipartite quantum systems, a profound understanding of the
relation and difference between the two GQDs is still awaited.

In this paper, we first analyze a class of symmetric $X$-shape states beyond the
Bell-diagonal form and show that there is a rigorous hierarchy relation between the two GQDs.
For multiqubit pure states, we show that the hierarchy relation is saturated and both GQDs are related
to the entanglement concurrence. Furthermore, we conduct a comparative investigation
on the two GQDs for a four-partite dynamical system consisting of two cavities interacting with
independent reservoirs. It is found that the GQD-2 can exhibit different sudden change
behaviors, while, in contrast to the GQD-2, the GQD-1 only evolves asymptotically, and is not
monogamous in both three- and four-qubit states. Finally, we discuss the monogamy property of
the square of GQD-1 in multipartite systems.

\section{The hierarchy and monogamy properties of two GQDs.}

According to the generalized Bloch representation, a two-qubit quantum state can be written
as \cite{nie00book}
\begin{equation}\label{4}
    \rho=\frac{1}{4}(\mbox{I}\otimes \mbox{I}+\sum_{i=1}^{3}x_i\sigma_i\otimes \mbox{I}
    +\sum_{i=1}^{3}y_i\mbox{I}\otimes \sigma_i+\sum_{i,j=1}^{3}\Gamma_{ij}\sigma_i\otimes
    \sigma_j),
\end{equation}
where $x_i=\mbox{Tr}\rho(\sigma_i\otimes \mbox{I})$, $y_i=\mbox{Tr}\rho(\mbox{I}\otimes \sigma_i)$
are components of local Bloch vectors with $\sigma_i$ being the Pauli matrix, and $\Gamma$
is the correlation matrix with element $\Gamma_{ij}=\mbox{Tr}\rho(\sigma_i\otimes \sigma_j)$.
In this case, the GQD-2 has an analytical formula \cite{dakic10prl,gir12prl}
\begin{equation}\label{5}
D_{g2}(\rho_{AB})=\frac{1}{2}(k_1+k_2+k_3-k_{max}),
\end{equation}
where $k_i$s are the eigenvalues of matrix $K=\vec{x}\vec{x}^{T}+\Gamma\Gamma^T$ with $k_{max}$
being the maximum.

So far, analytical results of GQD-1 have only been available for some specific classes of two-qubit
states, where a typical class is the $X$-shape states in the form
\begin{equation}\label{6}
\rho_{AB}^X=\left(
            \begin{array}{cccc}
              \rho_{11} & 0 & 0 & \rho_{41}^* \\
              0 & \rho_{22} & \rho_{32}^* & 0 \\
              0 & \rho_{32} & \rho_{33} & 0 \\
              \rho_{41} & 0 & 0 & \rho_{44} \\
            \end{array}
          \right).
\end{equation}
Ciccarello \emph{et al} proved the analytical formula of GQD-1 \cite{cic14njp}
\begin{equation}\label{7}
D_{g1}(\rho_{AB}^X)=\sqrt{\frac{\gamma_1^2\gamma_{max}-\gamma_2^2\gamma_{min}}
{\gamma_{max}-\gamma_{min}+\gamma_1^2-\gamma_2^2}},
\end{equation}
where $\gamma_1=2(\rho_{32}+\rho_{41})$, $\gamma_2=2(\rho_{32}-\rho_{41})$, and
$\gamma_3=1-2(\rho_{22}+\rho_{33})$ being the singular values of correlation matrix $\Gamma$, and
$\gamma_{min}=\mbox{min}\{\gamma_3^2,\gamma_1^2\}$ and
$\gamma_{max}=\mbox{max}\{\gamma_3^2,\gamma_2^2+x_3^2\}$ with $x_3$ being the $z$-component of
local Bloch vector.

We consider a class of symmetric two-qubit $X$-shape state $\rho_{AB}^{Xs}$, for which matrix
elements in Eq. (6) satisfy the conditions $\rho_{32}=0$ or $\rho_{41}=0$. For this class of
quantum states, we have the following hierarchy relation.

\emph{Theorem 1}. For symmetric $X$-shape states, two GQDs obey the following hierarchy relation
\begin{equation}\label{8}
D_{g1}^2(\rho_{AB}^{Xs})\geq D_{g2}(\rho_{AB}^{Xs}),
\end{equation}
where $D_{g1}=|\gamma_1|$ and $D_{g2}=(\gamma_1^2+\gamma_{min}^\prime)/2$ with
$\gamma_{min}^\prime=\mbox{min}\{\gamma_1^2, \gamma_3^2+x_3^2\}$, respectively.

\emph{Proof}. For $X$-shape states, the matrix $K$ of GQD-2 in Eq. (5) is diagonal and the eigenvalues
have the form $k_1=\gamma_1^2=4(\rho_{32}+\rho_{41})^2$, $k_2=\gamma_2^2=4(\rho_{32}-\rho_{41})^2$, and
$k_3=\gamma_3^2+x_3^2=[1-2(\rho_{22}+\rho_{33})]^2+[2(\rho_{11}+\rho_{22})-1]^2$. In the symmetric case
of $\rho_{32}=0$ or $\rho_{41}=0$, we further have $k_1=k_2=\gamma_1^2$
which results in $D_{g2}(\rho_{AB}^{Xs})=(\gamma_1^2+\gamma_{min}^\prime)/2$.
The GQD-1 in the symmetric $X$-shape states has the property $\gamma_1^2=\gamma_2^2$, and the measure
in Eq. (7) can be simplified to $D_{g1}=|\gamma_1|$ when the parameters $\gamma_{max}\neq\gamma_{min}$.
For the two parameters $\gamma_{max}$ and $\gamma_{min}$, there are three kinds of relations among the
singular values and local Bloch vector: i) $\gamma_3^2\geq\gamma_1^2$ and
$\gamma_3^2\geq\gamma_1^2+x_3^2$, ii) $\gamma_3^2\geq\gamma_1^2$ and $\gamma_3^2\leq\gamma_1^2+x_3^2$,
iii) $\gamma_3^2\leq\gamma_1^2$ and $\gamma_3^2\leq\gamma_1^2+x_3^2$, respectively. When
$\gamma_{max}=\gamma_{min}$, we can obtain $|\gamma_1|=|\gamma_2|=|\gamma_3|$ for all the three kinds
of relations, which corresponds to the homogeneous case and the GQD-1 has the form
$D_{g1}=|\gamma_1|$ \cite{cic14njp}. Therefore, for all symmetric $X$-shape states, the GQD-1 is
formulated as a unified form $D_{g1}(\rho^{Xs}_{AB})=|\gamma_1|$. Noting that the fact
$\gamma_1^2\geq\gamma_{min}^\prime$, we have the hierarchy relation in Eq. (8) and complete the proof.

In the pure state case, quantum correlation is usually equivalent to quantum entanglement. Here,
for $N$-qubit pure states, we have the following theorem.

\emph{Theorem 2}. In an arbitrary $N$-qubit pure state $\ket{\psi_{A_1A_2\cdots A_n}}$,
the GQDs under partition $A_1|A_2\cdots A_n$ are related to the entanglement quantified
by concurrence
\begin{eqnarray}\label{9}
&&D_{g1}(\psi_{A_1|A_2\cdots A_n})=C_{A_1|A_2\cdots A_n}(\psi)=2\sqrt{\lambda_0\lambda_1},\nonumber\\
&&D_{g2}(\psi_{A_1|A_2\cdots A_n})=C^2_{A_1|A_2\cdots A_n}(\psi)=4\lambda_0\lambda_1,
\end{eqnarray}
where the hierarchy relation is saturated with $\lambda_i$s being the eigenvalues of reduced density
matrix $\rho_{A_1}$.

\emph{Proof}. According to the Schmidt decomposition, an $N$-qubit pure state is equivalent to
a logical two-qubit state under the partition $A_1|A_2\cdots A_n$. Up to local unitary transformations,
the pure state has the form $\ket{\psi_{A_1A_2\cdots A_n}}=\sqrt{\lambda_0}\ket{0}_{A_1}
\ket{0}_{A_2\cdots A_n}+\sqrt{\lambda_1}\ket{1}_{A_1}\ket{1}_{A_2\cdots A_n}$, which belongs to the
class of symmetric $X$-shape states. Therefore, the GQD-1 for the $N$-qubit pure
state is $D_{g1}(\psi_{A_1|A_2\cdots A_n})=|\gamma_1|=2\sqrt{\lambda_0\lambda_1}$ in terms of theorem
$1$. On the other hand, the
concurrence for the logical two-qubit state is $C_{A_1|A_2\cdots A_n}(\psi)=2a_{00}a_{11}
=2\sqrt{\lambda_0\lambda_1}$ \cite{woo98prl}, leading to the first equation in the theorem. The
GQD-2 for the pure state is $D_{g2}(\psi_{A_1|A_2\cdots A_n})=2(1-\sum \lambda_i^2)$
\cite{luo11prl}. Utilizing the relation $1=(\lambda_0+\lambda_1)^2=2\lambda_0\lambda_1+\sum\lambda_i^2$,
we have $D_{g2}(\psi_{A_1|A_2\cdots A_n})=4\lambda_0\lambda_1=C^2_{A_1|A_2\cdots A_n}(\psi)$ which
is just the second equation in the theorem, and then the proof is completed.

Recently, Hu and Fan present a measure of measurement-induced nonlocality based on trace norm (MIN-1)
$N_1(\rho_{AB})=\mbox{max}_{\Pi^A}||\rho_{AB}-\Pi^A(\rho_{AB})||_1$ and the nonlocality in a
$2\otimes N$ pure state is $N_1(\psi)=2\sqrt{\lambda_0\lambda_1}$ \cite{mhu14arx}. According to
theorem 2, the GQD-1 is also equivalent to the MIN-1 in this case.

For mixed states, the GQDs are not equal to the entanglement in general, but the GQD-1 is
still related to the concurrence in symmetric $X$-shape states.

\emph{Lemma 1.} In the class of symmetric two-qubit $X$-shape states, the GQD-1 is not less than
the concurrence, i.e.,
\begin{equation}\label{10}
D_{g1}(\rho_{AB}^{Xs})\geq C(\rho_{AB}^{Xs}).
\end{equation}

\emph{Proof.} For the symmetric $X$-shape states, the concurrence can be expressed as
$C(\rho_{AB}^{Xs})=\mbox{max}\{0,|\gamma_1|-2\eta\}$ in which $\eta=\sqrt{\rho_{22}\rho_{33}}$
when the matrix element $\rho_{32}=0$ and $\eta=\sqrt{\rho_{11}\rho_{44}}$ when $\rho_{41}=0$.
According to theorem 1, we have $D_{g1}(\rho_{AB}^{Xs})=|\gamma_1|$ , from which
we can obtain the lemma after a direct comparison to the concurrence.

In many-body quantum systems, one of the most important properties is that entanglement is monogamous
\cite{ckw00pra}, and, as is known, this property can be used for constructing multipartite entanglement
measures \cite{ckw00pra,byw07pra,baw08pra}. For three-qubit pure states, the GQD-2 is monogamous
\cite{str12prl}, and this property is also satisfied by the square of quantum discord which results
in a genuine three-qubit quantum correlation measure \cite{bai13pra,bai14prl}. In comparison to the GQD-2, the
GQD-1 has a merit of contractility under local operations on the unmeasured party. Therefore,
it is natural to ask whether the GQD-1 is monogamous in the case of several qubits. Unfortunately,
it will be seen in the next section that the GQD-1 is not monogamous for the three-qubit pure states
and mixed states as well as the four-qubit pure states.

\section{The dynamical property of two GQDs in multipartite cavity-reservoir systems.}

Here we conduct a comparative study on GQD-1 and GQD-2 in a four-partite dynamical system consisting of
two cavities interacting with independent reservoirs, where the interaction Hamiltonian of a single
cavity-reservoir system is \cite{lop08prl,bel11ijqi}
\begin{equation}\label{11}
\hat{H}=\hbar \omega \hat{a}^{\dagger}\hat{a}+\hbar\sum_{k=1}^{N}\omega_{k}
\hat{b}_k^{\dagger}\hat{b}_k+\hbar\sum_{k=1}^{N}g_{k}(\hat{a}
\hat{b}_{k}^{\dagger}+\hat{b}_{k}\hat{a}^{\dagger}).
\end{equation}
The initial state is  $\ket{\Phi_0}=(\alpha\ket{00}+\beta\ket{11})_{c_1c_2}\ket{00}_{r_1r_2}$ in
which the dissipative reservoirs are in the
vacuum state. In the limit of $N\rightarrow \infty$ for a reservoir with a flat spectrum,
the output state of the multipartite system will be \cite{lop08prl}
\begin{equation}\label{12}
  \ket{\Psi_t}=\alpha\ket{0000}_{c_1r_1c_2r_2}+\beta\ket{\phi_t}_{c_1r_1}\ket{\phi_t}_{c_2r_2},
\end{equation}
where $\ket{\phi_t}=\xi(t)\ket{10}+\chi(t)\ket{01}$ with the amplitudes being
$\xi(t)=\mbox{exp}(-\kappa t/2)$ and $\chi(t)=[1-\mbox{exp}(-\kappa t)]^{1/2}$.

\begin{figure}
\begin{center}
\epsfig{figure=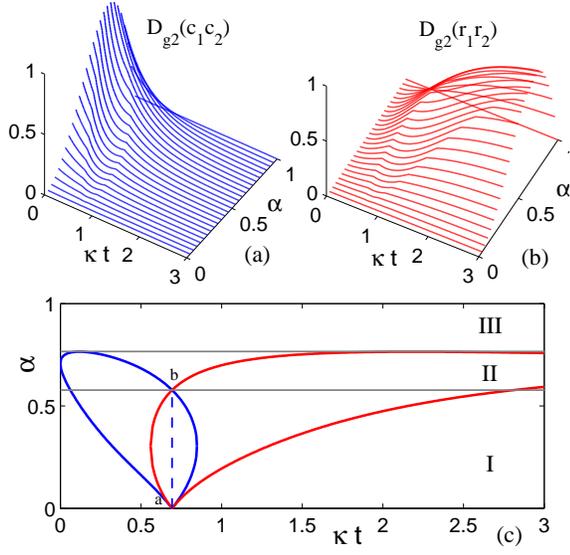,width=0.42 \textwidth}
\end{center}
\caption{(Color online) Dynamical evolution of GQD-2 as functions of the time
evolution $\kappa t$ and the initial state parameter $\alpha$: (a) $D_{g2}
(\rho_{c_1c_2})$, (b) $D_{g2}(\rho_{r_1r_2})$, (c) three types of correlation
evolution classified by the sudden change lines of cavity photons (blue line)
and reservoirs (red line).}
\end{figure}

In the dynamical evolution, the density matrix of two cavity photons does not preserve the Bell-diagonal
form ($x_3\neq 0$) and has the form
\begin{equation}\label{13}
\rho_{c_1c_2}=\left(
                \begin{array}{cccc}
                  \alpha^2+\beta^2\chi^4 & 0 & 0 & \alpha\beta\xi^2 \\
                  0 & \beta^2\xi^2\chi^2 & 0 & 0 \\
                  0 & 0 & \beta^2\xi^2\chi^2 & 0 \\
                  \alpha\beta\xi^2 & 0 & 0 & \beta^2\xi^4 \\
                \end{array}
              \right),
\end{equation}
which is just the \emph{symmetric} $X$-shape state with element $\rho_{32}=0$. We first consider the
correlation evolution of GQD-2 in the subsystem. According to theorem 1, we have
\begin{equation}\label{14}
D_{g2}(\rho_{c_1c_2})=[\gamma_1^2(\rho_{c_1c_2})+\gamma^\prime_{min}(\rho_{c_1c_2})]/2,
\end{equation}
where the parameters are $\gamma_1^2=4\alpha^2\beta^2\xi^4$ and
$\gamma^\prime_{min}=\mbox{min}\{\gamma_1^2,\gamma_3^2+x_3^2\}$ with
$\gamma_3^2=(1-4\beta^2\xi^2\chi^2)^2$ and $x_3^2=(1-2\beta^2\xi^2)^2$, respectively. For the subsystem
of two reservoirs, its density matrix is similar to that in Eq. (13) and the relation
$\rho_{r_1r_2}=S_{\xi\leftrightarrow\chi}[\rho_{c_1c_2}]$ is satisfied in which
$S_{\xi\leftrightarrow\chi}$ is a transformation exchanging the parameters $\xi$ and $\chi$.
Therefore, the GQD-2 of two reservoirs can be expressed as
\begin{equation}\label{15}
D_{g2}(\rho_{r_1r_2})=S_{\xi\leftrightarrow\chi}[D_{g2}(\rho_{c_1c_2})].
\end{equation}
In Fig.1(a) and Fig.1(b), $D_{g2}(\rho_{c_1c_2})$ and $D_{g2}(\rho_{r_1r_2})$ are plotted as
functions of the time evolution $\kappa t$ and the initial state amplitude $\alpha$, where we can
see that the sudden change behavior exists. Due to the intrinsic relation in Eq. (15), when the GQD-2
of two cavity photons experiences the sudden change, the same behavior must happen in the reservoir
systems, which is similar to the relation of entanglement sudden death and sudden birth
in the subsystems \cite{lop08prl,bai09pra,wen11epjd}.

For both cavity photons and reservoirs subsystems, the sudden change condition is
\begin{equation}\label{16}
\gamma_1^2=\gamma_3^2+x_3^2,
\end{equation}
and from which we plot the sudden change lines for $D_{g2}(\rho_{c_1c_2})$ (blue line) and
$D_{g2}(\rho_{r_1r_2})$ (red line) in Fig1.(c). As seen from the figure, there are three
types of correlation evolution. In type \emph{I} with
$\alpha\in(0,1/\sqrt{3})$, two $D_{g2}$s experience double sudden changes with a revival.
The revival occurs at the time $\kappa t=\mbox{ln}2$ and the blue dashed line $ab$ indicates the
minimum of $D_{g2}$s in the overlap area marked out by the sudden change lines. In type
\emph{II} with $\alpha\in (1/\sqrt{3}, 0.7647)$, although the revival
behavior disappears, the double sudden changes still exist except for the case
$\alpha=1/\sqrt{2}$ exhibiting single sudden change behavior. In type \emph{III}
with $\alpha\geq 0.7647$, two quantum correlations evolve asymptotically, and the sudden change
behavior disappears. Similarly, for the subsystems $c_1r_1$ and $c_1r_2$, we can also use the
sudden change condition in Eq. (16) to determine the type of correlation evolution.

\begin{figure}
\begin{center}
\epsfig{figure=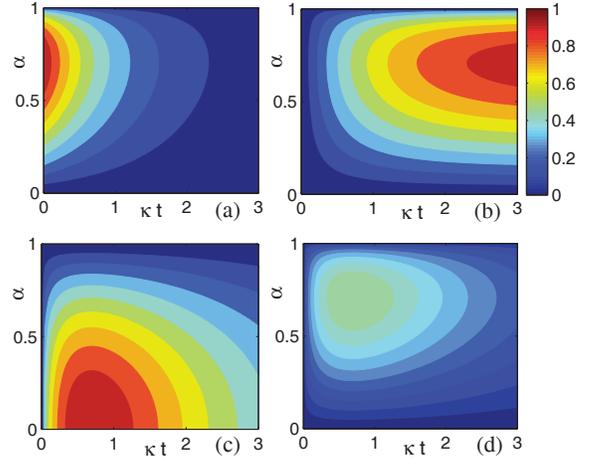,width=0.42 \textwidth}
\end{center}
\caption{(Color online) Asymptotical evolution of GQD-1 in different subsystems
as functions of the time evolution $\kappa t$ and the initial state amplitude
$\alpha$: (a) $D_{g1}(\rho_{c_1c_2})$, (b) $D_{g1}(\rho_{r_1r_2})$, (c)
$D_{g1}(\rho_{c_1r_1})$, (d) $D_{g1}(\rho_{c_1r_2})$.}
\end{figure}

We now analyze two-qubit GQD-1 in multipartite cavity-reservoir systems, in which all the
two-qubit states are in the symmetric $X$-shape form. Based on theorem 1, the correlation
equates to the first singular value $|\gamma_1|$ and we have
\begin{eqnarray}\label{17}
&&D_{g1}(\rho_{c_1c_2})=2\alpha\beta\xi^2,
D_{g1}(\rho_{r_1r_2})=2\alpha\beta\chi^2,\nonumber\\
&&D_{g1}(\rho_{c_1r_1})=2\beta^2\xi\chi,
D_{g1}(\rho_{c_1r_2})=2\alpha\beta\xi\chi,
\end{eqnarray}
respectively. In Fig.2, the four $D_{g1}$s are plotted as functions of parameters $\kappa t$
and $\alpha$, where all the quantum correlations evolve asymptotically.
Via the hierarchy relation in theorem 1, the GQD-1 is related to the GQD-2.
For example,
$D_{g1}^2(\rho_{c_1c_2})=D_{g2}(\rho_{c_1c_2})$ when the sudden change of $D_{g2}(\rho_{c_1c_2})$ does
not occur, and when the sudden change appears the relation is strictly
$D_{g1}^2(\rho_{c_1c_2})>D_{g2}(\rho_{c_1c_2})$ in the sudden change area. The case for other
subsystems is similar.

\begin{figure}
\begin{center}
\epsfig{figure=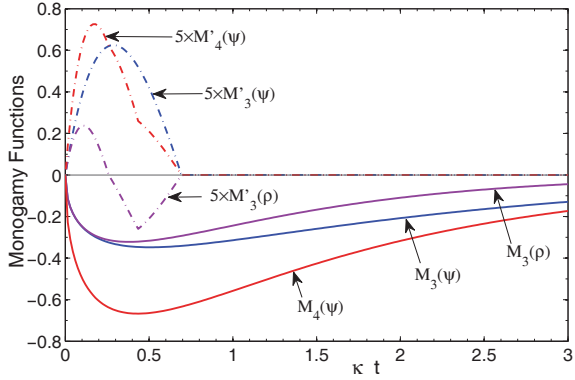,width=0.42 \textwidth}
\end{center}
\caption{(Color online) Monogamy analysis of GDD-1 and GQD-2 in multipartite
cavity-reservoir systems, where $M_3(\Psi)$ (blue solid line), $M_3(\rho)$
(purple solid line), $M_4(\Psi)$ (red solid line), and $M_3^\prime(\rho)$
(purple dash-dotted line) are not monogamous but $M_3^\prime(\Psi)$ (blue
dash-dotted line) and $M_4^\prime(\Psi)$ (red dash-dotted line) are monogamous.}
\end{figure}

The GQD-2 was indicated to be monogamous in three-qubit pure states~\cite{str12prl}.
Whether the GQD-1 is still monogamous in these states? To answer this question, we turn to
analyze the monogamy property of the two GQDs in multipartite cavity-reservoir systems.

In the output state of Eq. (12), subsystem $c_2r_2$ can be regarded as a logic qubit, and then the
reduced state $\rho_{c1(c2r2)}$ is equivalent to a two-qubit state which is in the symmetric
$X$-shape form and its GQD-1 is $D_{g1}(\rho_{c_1(c_2r_2)})=2\alpha\beta\xi$.
In multipartite cavity-reservoir systems, the monogamy property of three-qubit pure states can
be checked via the quantity $M_3(\ket{\Psi_t})=D_{g1}(\ket{\Psi}_{c_1|r_1(c_2r_2)})
-D_{g1}(\rho_{c_1r_1})-D_{g1}(\rho_{c_1|(c_2r_2)})$ in which the multiqubit GQD-1 can be calculated by
theorem 2. Similarly, for the cases of three-qubit mixed states and four-qubit pure states, we can
utilize $M_3(\rho_t)=D_{g1}(\rho_{c_1(c_2r_2)})-D_{g1}(\rho_{c_1c_2})-D_{g1}(\rho_{c_1r_2})$ and
$M_4(\ket{\Psi_t})=D_{g1}(\ket{\Psi}_{c_1|r_1c_2r_2})-D_{g1}(\rho_{c_1r_1})-D_{g1}(\rho_{c_1c_2})
-D_{g1}(\rho_{c_1r_2})$, respectively. In the same way, we can define the quantities
$M_3^\prime(\ket{\Psi_t})$, $M_3^\prime(\rho_t)$, and $M_4^\prime(\ket{\Psi_t})$ with the corresponding
$D_{g2}$s.
In Fig.3, these monogamy quantities are plotted as functions of the time evolution
$\kappa t$, where the
initial state amplitude is chosen to be $\alpha=1/\sqrt{2}$. The function $M_3^\prime(\ket{\Psi_t})$
(blue dash-dotted line) is always nonnegative, which coincides with the fact that the GQD-2 is
monogamous in three-qubit pure states \cite{str12prl}. In three-qubit mixed states,
$M_3^\prime(\rho_t)$ can be positive or negative (purple dash-dotted line), and then it is not
monogamous. In four-qubit pure states, although $M_4^\prime(\ket{\Psi_t})$ (red dash-dotted line) is
monogamous, but the property in general case cannot be guaranteed. For the GQD-1 in multipartite
cavity-reservoir systems, we find all the functions $M_3(\ket{\Psi_t})$ (blue solid line),
$M_3(\rho_t)$ (purple solid line), and $M_4(\ket{\Psi_t})$ (red solid line) are negative, which
indicates the GQD-1 is not monogamous.

\section{Discussion and conclusion.}

Although the GQD-1 itself is not monogamous in the cavity-reservoir system, it is noted that the
square of
GQD-1 may be monogamous in this multipartite system. After some derivations,
we can obtain $M_3^{(2)}(\ket{\Psi_t})=D_{g1}^2(\ket{\Psi}_{c_1|r_1(c_2r_2)})-D_{g1}^2(\rho_{c_1r_1})
-D_{g1}^2(\rho_{c_1|(c_2r_2)})=0$ for all the values of parameters $\kappa t$ and
$\alpha$. In addition,  quantified by the square of GQD-1, we can get
$M_3^{(2)}(\rho_t)=0$ and $M_4^{(2)}(\ket{\Psi_t})=0$ as well.
Moreover, it is also interesting to look into the monogamy property in a nontrivial $N$-qubit case \cite{note1}, i.e., a generalized $W$ state
\begin{equation}\label{18}
\ket{W_N}=a_1\ket{10\cdots 0}+a_2\ket{01\cdots 0}+\cdots +a_n\ket{00\cdots 1}.
\end{equation}
Its two-qubit reduced density matrix $\rho_{1j}$ is in the symmetric $X$-shape.
According to lemma 1, it is  known that the GQD-1 is not less
than the concurrence, and thus
$D_{g1}(\rho_{1j})=C(\rho_{1j})=2a_1a_j$ in this case. Combining this property with theorem 2, we obtain
$M_N^{(2)}(\ket{W})=C^2(\ket{W}_{A_1|A_2\cdots A_n})-\sum_{j\neq1} C^2(\rho_{A_1A_j})=0$, which implies
that the square of GQD-1 is monogamous \cite{note2}. It is worthwhile pointing out that  the monogamy property of the
square of GQD-1 even in a general three-qubit (or four-qubit pure)  state is still an open problem
mainly due to the hardness of computing the GQD-1.

In conclusion, we have made a comparative study on two GQDs in multipartite systems. In symmetric
$X$-shape states, there exists a hierarchy relation between the GQDs, and the GQD-1 has a simple
expression without maximization and minimization. Moreover, for multiqubit pure states, the hierarchy
relation is saturated and the GQDs are related to the entanglement concurrence. Furthermore, we have
analyzed the dynamical property of two GQDs in multipartite cavity-reservoir systems,
and found that (i) the GQD-2 can exhibit various sudden change behaviors while the GQD-1 only evolves
asymptotically; and (ii) in contrast to the GQD-2 being monogamous in three-qubit pure
states, the GQD-1 is not monogamous in both three- and four-qubit states.

\emph{Acknowledgments.} --
This work was supported by the RGC of Hong Kong under Grant No. HKU7058/11P.
Y.K.B. and T.T.Z. were also supported by NSF-China (under Grant No. 10905016) and Hebei
NSF (under Grant No. A2012205062).


\begin{thebibliography}{99}
\bibitem{note1}         For the trivial case of generalized $N$-qubit $GHZ$ state
                        $\ket{G_N}=a\ket{00\cdots 0}+b\ket{11\cdots 1}$, since all its
                        two-qubit reduced states are classical states, the monogamy score of the square
                        of GQD-1 is just $D_{g1}^2({A_1|A_2\cdots A_n})$ indicating the monogamy
                        property. The case for GQD-2 is similar.
\bibitem{note2}         According to the hierachy relation
                        $D_{g1}^2(\rho_{A_1A_j}^{Xs})\geq D_{g2}(\rho_{A_1A_j}^{Xs})$ in theorem 1, we
                        can further obtain that the GQD-2 is also monogamous in the generalized
                        $N$-qubit $W$ state.
\end{thebibliography}
\end{document}